\documentclass{bmcart}

\usepackage[
type={CC},
modifier={by-sa},
version={3.0},
]{doclicense}

\usepackage[utf8]{inputenc} 
\usepackage{amssymb}
\usepackage{float}

\usepackage{graphicx,slashed,color,xspace,colortbl,url}
\usepackage{mathptmx,mathtools}
\usepackage{setspace}
\usepackage{xkeyval}

\startlocaldefs
\endlocaldefs

\begin{document}


\begin{figure}[h!]
\centering
\includegraphics[keepaspectratio=true]{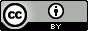}

Comparison Between IPv4 to IPv6 Transition Techniques by Edwin Cordeiro is licensed under a Creative Commons Attribution 4.0 International License. 
\end{figure}

\begin{frontmatter}

\begin{fmbox}
\dochead{Experience Report}


\title{Comparison Between IPv4 to IPv6 Transition Techniques}


\author[
   addressref={aff1},                   
   corref={aff1},                       
   noteref={n1},                        
   email={edwin@scordeiro.net}   
]{\inits{ES}\fnm{Edwin S} \snm{Cordeiro}}
\author[
   addressref={aff2},
   email={rodrigo.carnier@usp.br}
]{\inits{RM}\fnm{Rodrigo M} \snm{Carnier}}
\author[
   addressref={aff2},
   email={wzucchi@lps.usp.br}
]{\inits{WL}\fnm{Wagner L} \snm{Zucchi}}


\address[id=aff1]{
  \orgname{Institute for Technological Research, São Paulo}, 
  \street{Av. Prof. Almeida Prado 532, Cid. Universitária - Butantã},                     %
  \postcode{05508-901},                                
  \city{São Paulo},                              
  \cny{Brazil}                                    
}
\address[id=aff2]{%
  \orgname{University of São Paulo, USP},
  \street{Av. Prof. Luciano Gualberto, travessa 3 nº 380, Cid. Universitária - Butantã},
  \postcode{05508-010}
  \city{São Paulo},                              
  \cny{Brazil},                                    
}


\begin{artnotes}
\note[id=n1]{Principal corresponding author} 
\end{artnotes}

\end{fmbox}


\begin{abstractbox}

\begin{abstract} 

The IPv4 addresses exhaustion demands a protocol transition from IPv4 to IPv6. The original transition technique, the dual stack, is not widely deployed yet and it demanded the creation of new transition techniques to extend the transition period.

This work makes an experimental comparison of techniques that use dual stack with a limited IPv4 address. This limited address might be a RFC 1918 address with a NAT at the Internet Service Provider (ISP) gateway, also known as Carrier Grade NAT (CGN), or an Address Plus Port (A+P) shared IPv4 address. The chosen techniques also consider an IPv6 only ISP network. The transport of the IPv4 packets through the IPv6 only networks may use IPv4 packets encapsulated on IPv6 packets or a double translation, by making one IPv4 to IPv6 translation to enter the IPv6 only network and one IPv6 to IPv4 translation to return to the IPv4 network. The chosen techniques were DS-Lite, 464XLAT, MAP-E and MAP-T.

The first part of the test is to check some of the most common usages of the Internet by a home user and the impacts of the transition techniques on the user experience. The second part is a measured comparison considering bandwidth, jitter and latency introduced by the techniques and processor usage on the network equipment. 
%
\end{abstract}




\end{abstractbox}
%

\end{frontmatter}


\section{Introduction}
\label{introduction} 

Vinton Cerf was asked in an interview to Slashdot what he would change about TCP/IP if he could travel back in time. His reply was: "I wish I had realized we'd need more than 32 bits of address space! At the time, I thought this was still an experiment and that, if successful, we would develop a production version. I guess IPv6 is the production version!" \cite{Cerf}.

The IPv6 main change is the 128 bits addresses, but it also added new functionalists like the neighbor discovery protocol and router advertisement. It also introduced a simpler fixed size header and extension headers for additional information that aren't needed for packet routing.

The transition from the IPv4 to the IPv6 was planned to avoid problems on the currently working IPv4 networks. This would be accomplished by creating dual stack networks and, when all the devices had IPv6 addresses, the IPv4 would become obsolete. Unfortunately this plan wasn't fully executed yet and the IPv4 addresses are ending before the IPv6 arrives to all devices on the Internet \cite{Huston}. As the IPv4 and the IPv6 are not compatible, other transition techniques were created on IETF Softwires group \cite{softwire-mail}.

This work provides an experimental comparison of existing techniques that transport IPv4 packets through IPv6 only networks.

\section{Related Work}
\label{related_work}

Some of the first transition techniques considered that the Internet still were mainly IPv4 and because of that the created techniques tried to enable IPv6 communication through IPv4 networks, like 6to4 \cite{RFC3056} and 6rd \cite{RFC5969}. Currently the techniques are being developed to do the opposite, to enable IPv4 communication through IPv6 only networks. 

The first option to achieve this is by tunneling IPv4 packets through the IPv6 network, using encapsulated IPv4 packets as data on IPv6 packets. The second option is a  double translation, in which IPv4 packets are translated to IPv6 packets as specified in RFC6145 \cite{RFC6145} and on the edge of the ISP network a second translation from IPv6 back to IPv4 is made. 

Another problem faced for those techniques is the lack of IPv4 addresses, so they have considered ways to reduce the need of them, by sharing IPv4 using NAT at the ISP network (CGN) \cite{d-nat444} or by using the source port as an extension of the address, A+P (Address Plus Port Sharing) \cite{MBC+08} restricting the ports that each customer can use.

Considering those options, some techniques were created and others are still being proposed. The four techniques compared in this article are examples of possible combination of the described techniques for an IPv6 only ISP network with shared IPv4 adresses: DS-Lite \cite{RFC6333}, MAP-E \cite{d-map-e} e MAP-T \cite{d-map-t}, 464XLAT \cite{d-464xlat}. The basic comparison of the techniques is available on table \ref{func_comp}.

\begin{table}[h!]
\caption{Comparison of IPv6 only ISP transition techniques}
\begin{tabular}{ccccc}
 & Tunneling & \shortstack{\\Double\\ Translation} & CGN & A+P \\
\hline
\hline DS-Lite & X & & X & \\
\hline MAP-E   & X & & & X \\
\hline 464XLAT & & X & X & \\
\hline MAP-T   & & X & & X \\
\hline
\end{tabular} 
\label{func_comp}
\end{table}


\section{Network Topology}
\label{network_topology}

\begin{figure}[h!]
\centering
\includegraphics[width=.9\linewidth, keepaspectratio=true]{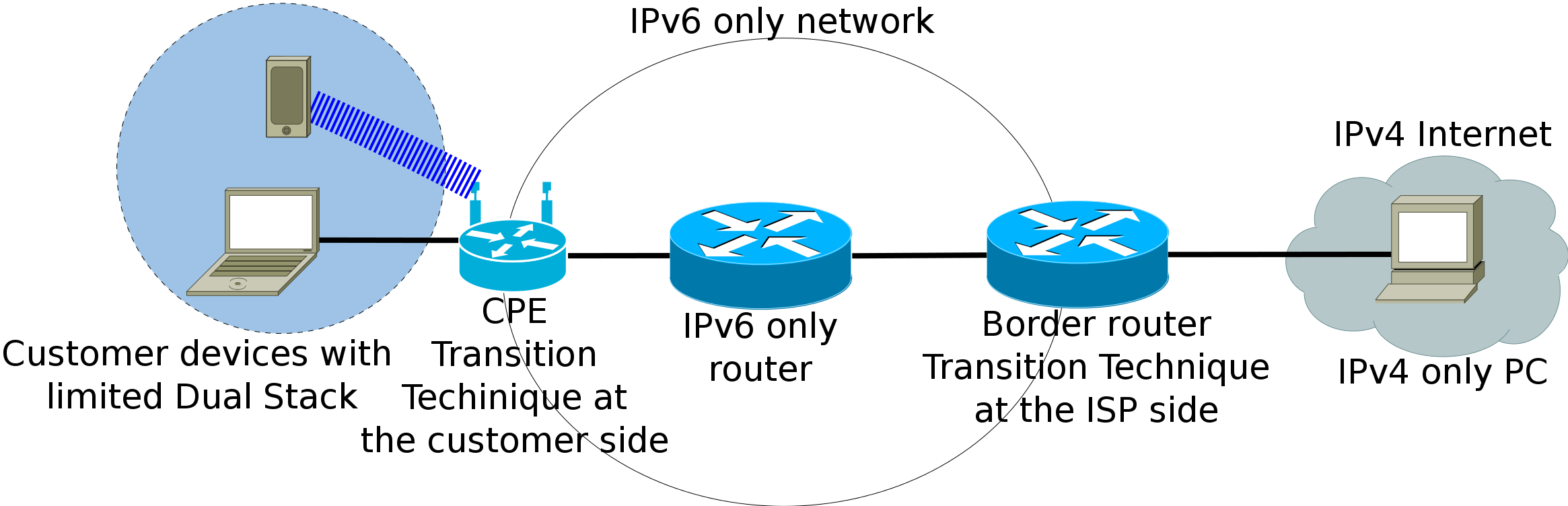}
\caption[Test network topology] {Test network topology\label{test-topology}}
\end{figure}

The network topology is the same for all the tested techniques and it is explained on figure \ref{test-topology} and includes a customer device to access Internet, a customer premises equipment (CPE) implementing the transition technique on the customer side, an IPv6 only router emulating an IPv6 only ISP network and a provider edge router implementing the ISP part of the transition technique. The CPE and the IPv6 only router are a TP-Link 1043ND Wireless Router with OpenWRT \cite{openwrt} firmware and the provider edge router is a Linux desktop with Ubuntu or Fedora (depending on the technique).

The software used to implement the DS-Lite network was developed by the Internet Systems Consortium \cite{dslite}, a public benefit corporation dedicated to supporting the infrastructure of the Internet by developing software, protocols, and operations. The DS-Lite, the oldest and most mature of the analyzed techniques, is implemented on a lot of network equipment from the most important manufacturers.  

MAP-E and MAP-T implementation \cite{map} were developed by China Education and Research Network (CERNET) as proof of concept for the protocol they were proposing at the IETF. Originally MAP-E and MAP-T were only different operating modes for a single protocol called MAP, because of that MAP-E and MAP-T share the same source code. Later IETF decided MAP should be divided in two, that MAP-E should proceed in the standard track and MAP-T should proceed in the experimental track. Both MAP-E or MAP-T are still IETF's drafts and still aren't implemented in large scale.   

Despite being approved in the IETF as an informational document and currently in use at the T-Mobile USA's network \cite{tmobile} the 464XLAT is currently only supported at Android devices \cite{464xlatdep}. As 464XLAT doesn't have a Linux compatible open source implementation for the CPE, like DS-Lite, MAP-E and MAP-T, it was used Tayga \cite{tayga}, a NAT64 \cite{RFC6146} software, to implement the CPE's IPv4 to IPv6 translation. Tayga was also used in its original function as NAT64 Border Router for IPv6 to IPv4 translations needed to implement the 464XLAT. The last Tayga update was on June 2011 and as the software isn't updated in a long time it wasn't optimized for small routers nor to the CPE's IPv4 to IPv6 translation.

\section{Comparison Results}
\label{comparison_results}

The comparison was made through a qualitative analysis, testing for noticeable problems or performance issues on common applications and uses of the Internet, and through a quantitative analysis that considered measurement results on networks that implemented the selected techniques.

\subsection{Qualitative Analysis}
\label{qualitative_analysis}

Ten of the most used applications and services \cite{CETIC, Pew} were chosen to be tested on networks running DS-Lite, MAP-E, 464XLAT and MAP-T: Internet search, Email access (webmail and smtp), Social networks, Video streaming, Peer-to-peer download (bit torrent), News websites, Internet home banking, Video call, Online gaming and L2TP / IPsec VPN.

Those applications were tested to check if they worked correctly and if any performance or functionality issue could be noticed by the end user. The tests were made on desktops with Windows 7, Ubuntu Linux 12.04 and MAC OS 10.8.3. Mobile devices were also tested with Android 4.1.2 and Apple IOS 6.1.3.

The peer-to-peer download using the bit torrent protocol failed on all techniques, because none of then allowed incoming connections for the devices. Despite being able to download the target file on all techniques it failed to receive incoming connections after the download of the file is finished, so the client fails to work as a seed which is a fundamental part of the protocol.

The L2TP / IPsec VPN also had very bad results as it failed to work on all operating systems when running through MAP-E, 464XLAT and MAP-T networks. When running on the DS-Lite network it worked correctly on Windows 7 and Android, totally failed on MAC OS and Apple IOS and had a strange result on Linux as the VPN session was successfully established but no traffic was being sent or received through the VPN.

The online game used on the testing was League of Legends \cite{lol} so it was only tested on Windows 7 and MAC OS. The game worked fine on the DS-Lite and 464XLAT networks for both operating systems. On the networks with MAP-E and MAP-T it worked fine at MAC OS version, but the Windows version constantly lost connection with the game servers. No significant delay was noticed during the gameplay.   

The Skype software \cite{skype} was used for the video call and it clearly suffered a quality loss when running on all the tested networks and all operating systems. The video call suffered a quality loss that was noticeable on the transmitted and received videos that became pixelated and blurry. The video call was particularly worse on Android at all the tested networks, but all the operating systems suffered some quality loss on all tested networks. The following figures \ref{skype_windows} and \ref{skype_linux} are samples acquired to demonstrate the differences. On the Windows sample (figure \ref{skype_windows}) the difference is very easy to notice as the quality is really bad on the 464XLAT network, showing a very blured image. On the Linux sample (figure \ref{skype_linux}) the difference is still perceptible, but it isn't so bad on the DS-Lite test network as it were on the 464XLAT test network.

\begin{figure}[h!]
  \centering
  \includegraphics[width=.9\linewidth , keepaspectratio=true]{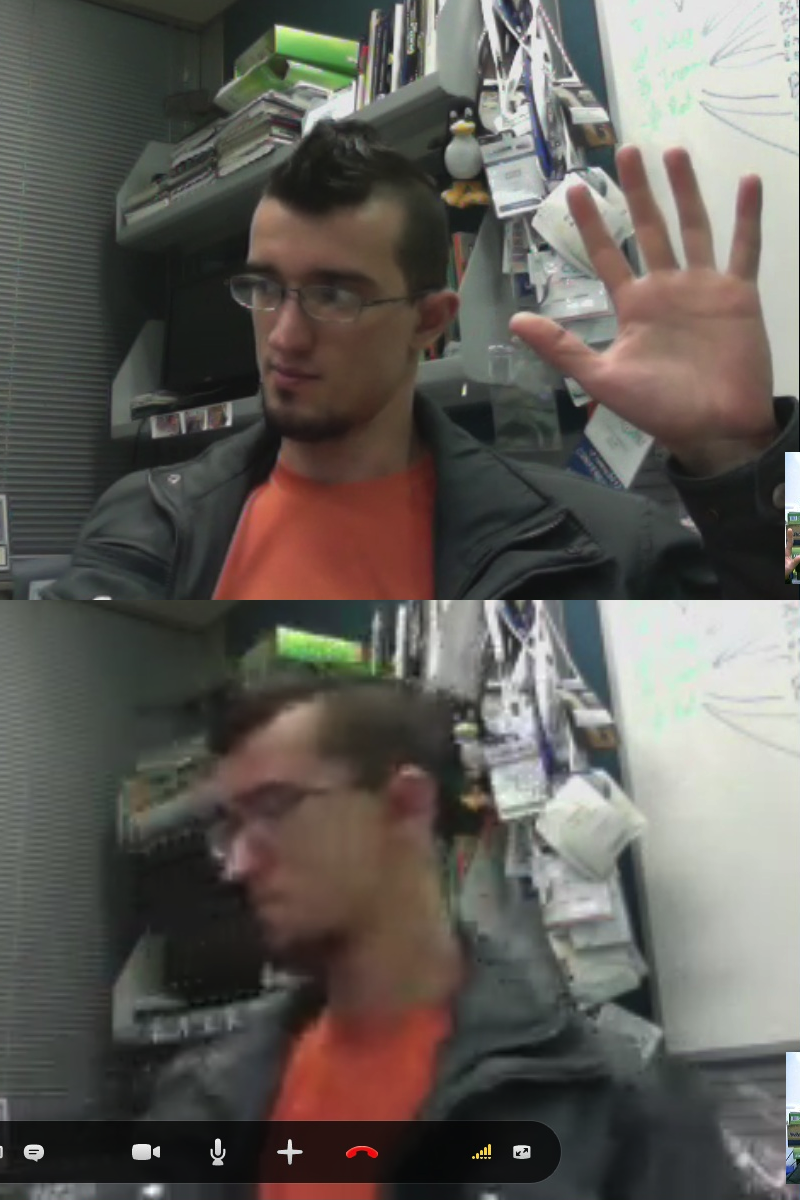}
  \caption[Skype on Windows on a network with native IPv4 and on the 464XLAT network]
   {Skype on Windows on a network with native IPv4 and on the 464XLAT network\label{skype_windows}}

\end{figure} 

\begin{figure}[h!]
  \centering
  \includegraphics[width=.9\linewidth, keepaspectratio=true]{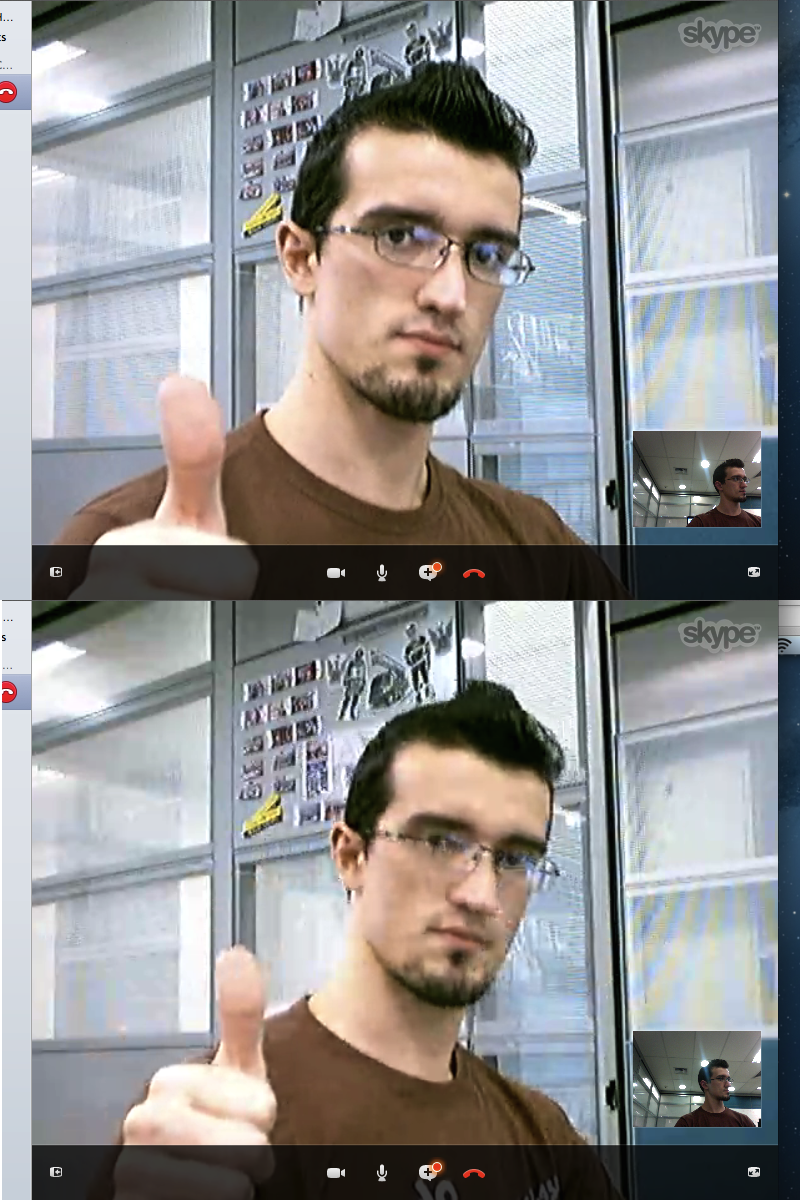}
  \caption[Skype on Linux on a network with native IPv4 and on the DS-Lite network]
   {Skype on Linux on a network with native IPv4 and on the DS-Lite network\label{skype_linux}}
\end{figure} 

The rest of the tested services (Internet search, Email access (webmail and smtp), Social networks, Video streaming, News websites and Internet home banking) had no noticeable problem.

\subsection{Quantitative Analysis}
\label{quantitative_analysis}

The software iperf \cite{iperf} was used to generate network traffic and to measure network quality parameters like jitter, packet loss and throughput. ICMP Echo Request and Echo Reply with different packet sizes were used to measure latency. Others parameters were measured, using sar \cite{sar}, to help to determinate the technique impact on network performance through the measurement of the processor usage on the network equipment running the transition technique.

On those performance tests the size of the packets were also a variable using packets of 100, 500, 1000 and 1400 bytes and the throughput was measured using TCP and UDP. On the UDP test the iperf was configured to send 100 Mbps and if the routers were unable to transmit all the packets it would cause some packet loss and result on a smaller effective speed.

\begin{figure}[h!]
  \centering
  \includegraphics[width=.9\linewidth, keepaspectratio=true]{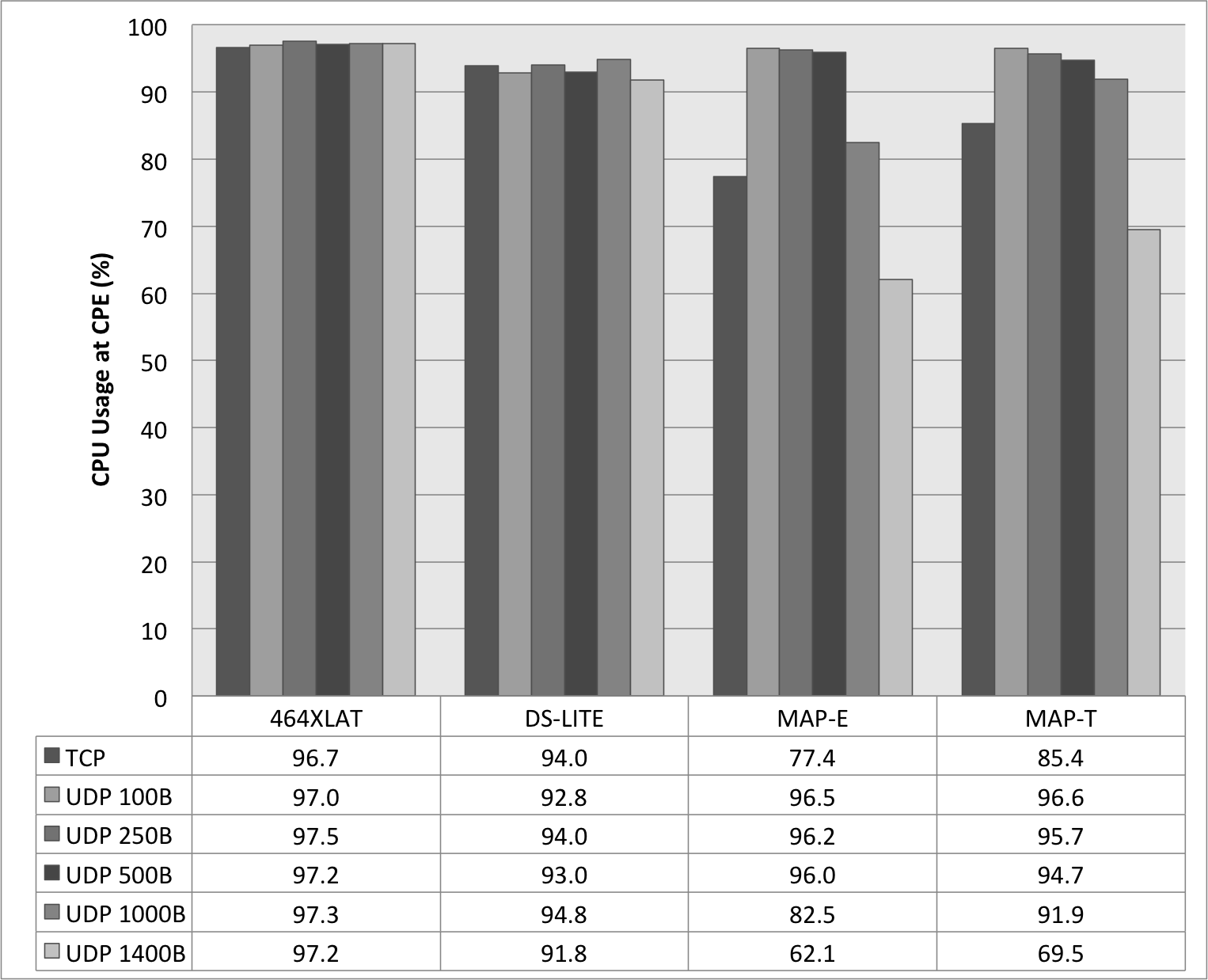}
  \caption[CPU Usage x Packet Size x Technique] {CPU Usage x Packet Size x Technique\label{cXtpXt}}
\end{figure}

Analyzing the CPU usage at the CPE's running the transition technique it was possible to evaluate the impact the of the technique in the CPE, as presented on the figure \ref{cXtpXt}. The results demonstrated that the CPE's CPU were completed used on all 464XLAT and DS-Lite tests. In the MAP-E and MAP-T networks the CPU was at maximum capacity for the 100, 250 and 500 bytes UDP packets, but for the TCP and the 1000 and 1400 bytes UDP tests a decrease on CPU usage was noticeable. This result was considered unexpected because neither 464XLAT or DS-Lite make stateful translations on the CPE, they make respectively a stateless IPv4 to IPv6 translation and encapsulation and desencapsulation of IPv4 packets to and from IPv6 packets. 

\begin{figure}[h!]
  \centering
  \includegraphics[width=.9\linewidth, keepaspectratio=true]{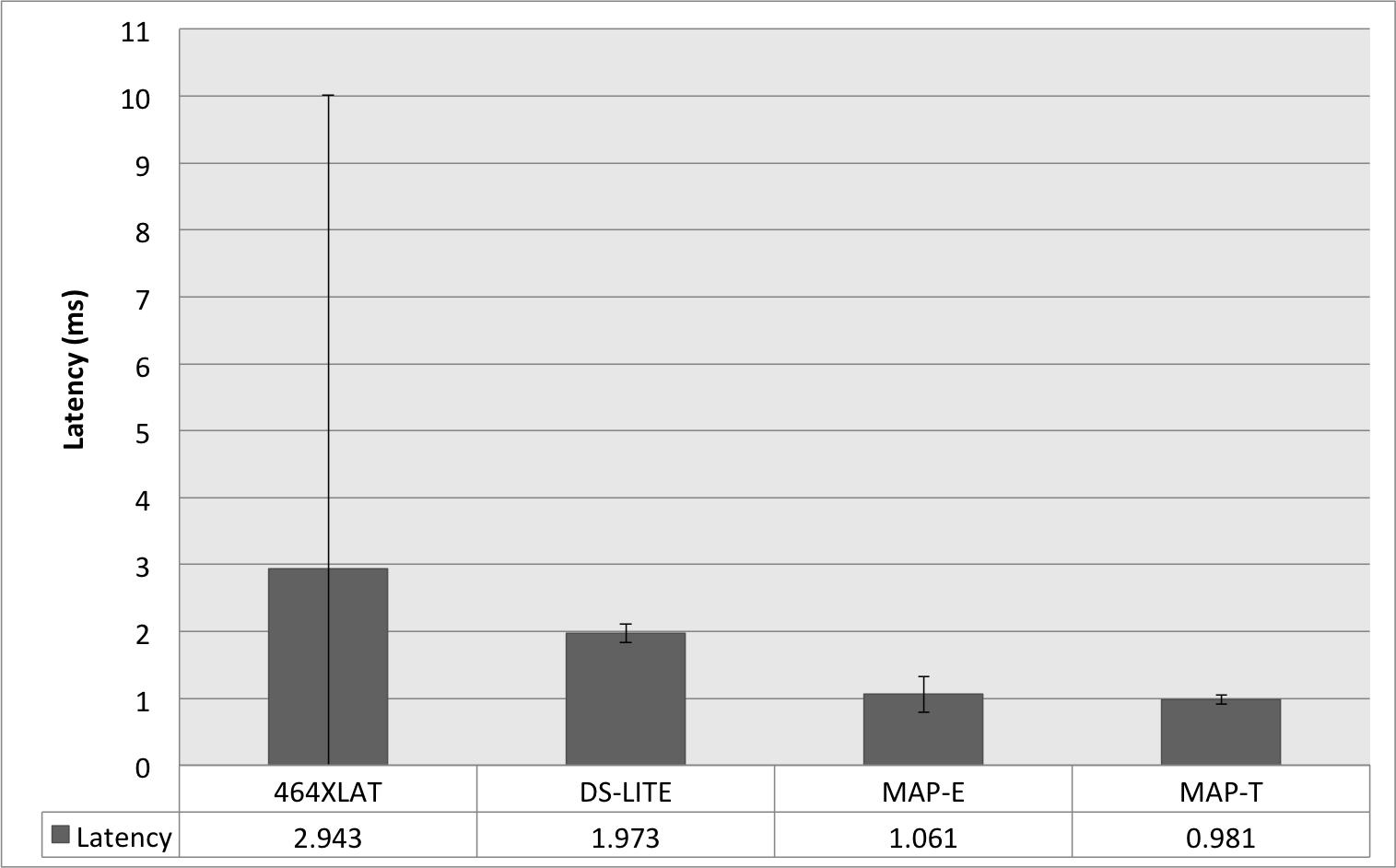}
  \caption[Latency x Technique] {Latency x Technique\label{apXt}}
\end{figure}

The other analyzed parameters results seemed to be related with the CPU usage in the CPE, and with lower CPU usage better results appeared. The latency results, presented on figure \ref{apXt}, shows that 464XLAT had a latency three times bigger than MAP-E and MAP-T that where less CPU intensive. DS-Lite had a intermediate result being 1 ms faster than 464XLAT and 1 ms slower than MAP-E and MAP-T, this result being compatible with the CPU usage that was smaller than 464XLAT, but bigger than MAP-E and MAP-T.

\begin{figure}[h!]
  \centering
  \includegraphics[width=.9\linewidth, keepaspectratio=true]{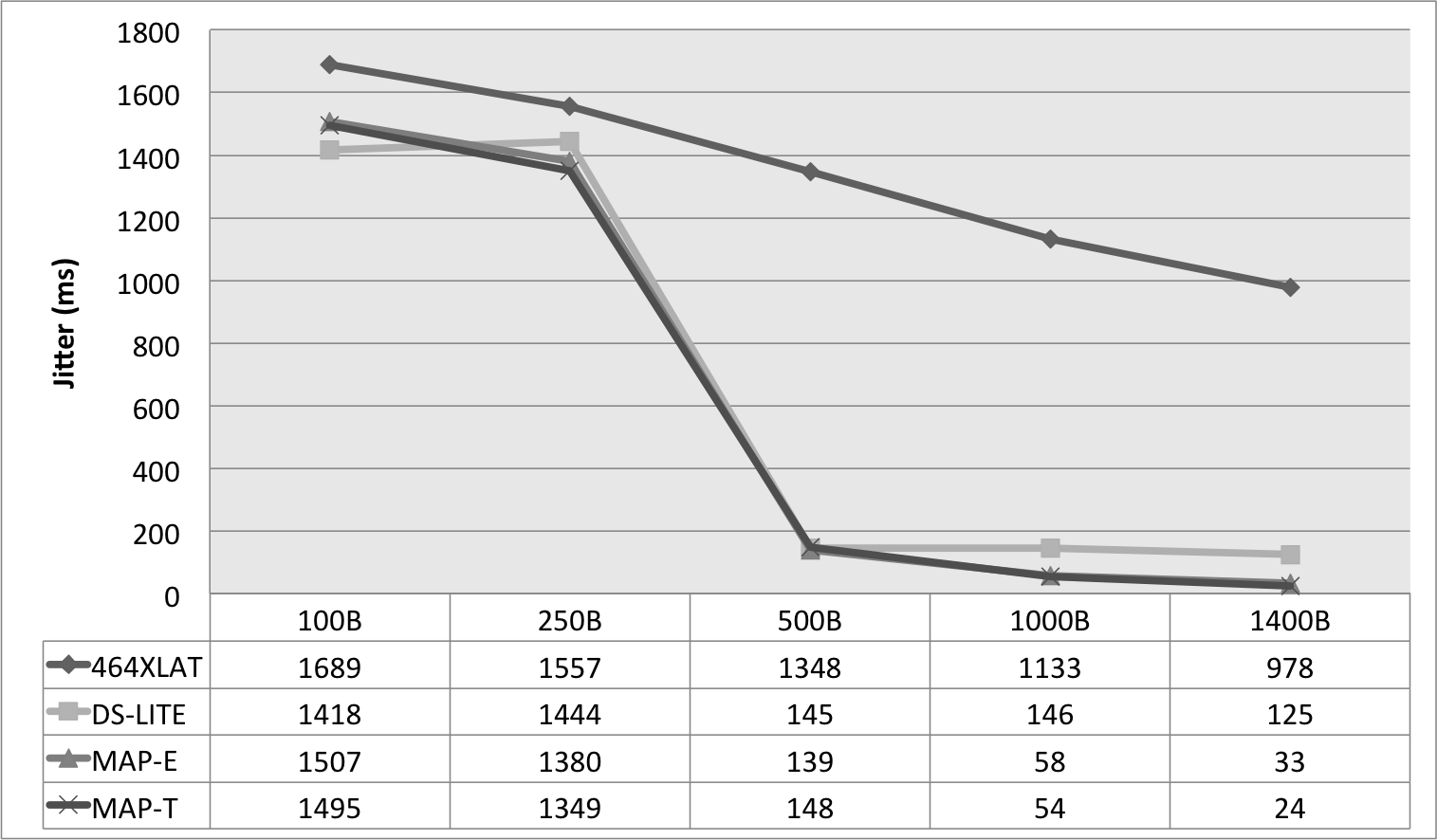}
  \caption[Jitter x Packet Size x Technique]{Jitter x Packet Size x Technique\label{jXtpXt}}
\end{figure}

Analyzing the jitter tests MAP-E, MAP-T and DS-Lite had similar results for the cases with high CPU usage. They had a very bad result for 100 and 250 bytes packets, when the CPU usage was very high. On the 500 bytes packets a considerable better result was noticeable, but while DS-Lite jitter stabilized at bigger than 100 ms values, MAP-E and MAP-T continued to have better results with bigger packets. 464XLAT was the worst of all in this test and besides the improvement for bigger packets the almost 1000 ms of its best result, still is a very bad jitter for a network. The results are presented on figure \ref{jXtpXt}.

\begin{figure}[h!]
  \centering
  \includegraphics[width=.9\linewidth, keepaspectratio=true]{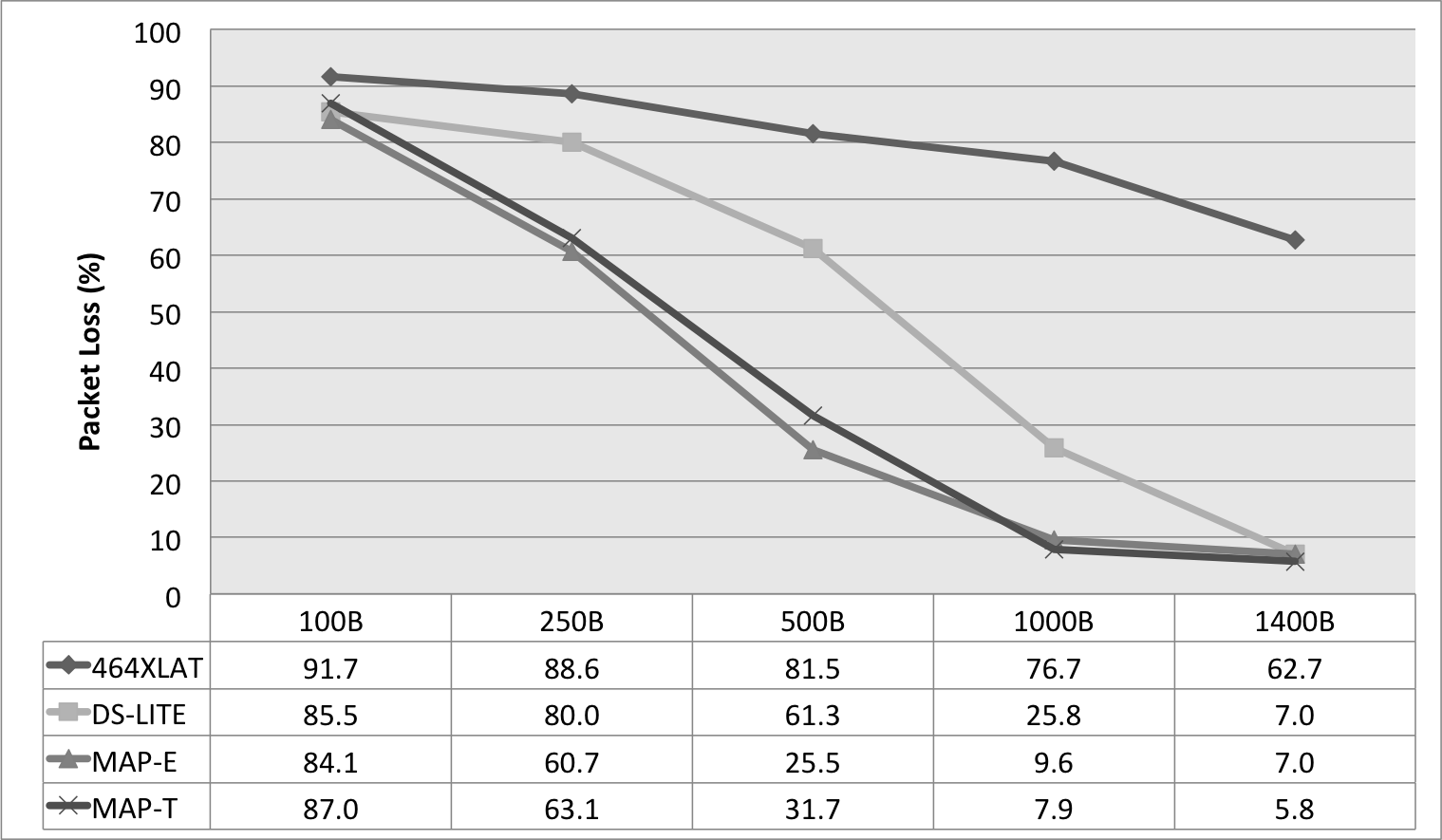}
  \caption[Packet loss x Packet Size x Technique]{Packet loss x Packet Size x Technique\label{ppXtpXt}}
\end{figure}

Looking at the packet loss no technique was able to provide an acceptable result for small packets as showed on figure \ref{ppXtpXt}. For the 100 bytes packets the best result was at MAP-E test with 84\% of packets lost and the worst was 464XLAT with more than 91\% of packets lost. MAP-E, MAP-T and DS-Lite had significant improvement for the bigger packets, but despite similar results for the 1400 bytes packets, DS-Lite had worse results for the intermediate sizes. 464XLAT had an improvement for big packets, but even with big packets it had more than 60\% of packet lost.

\begin{figure}[h!]
  \centering
  \includegraphics[width=.9\linewidth, keepaspectratio=true]{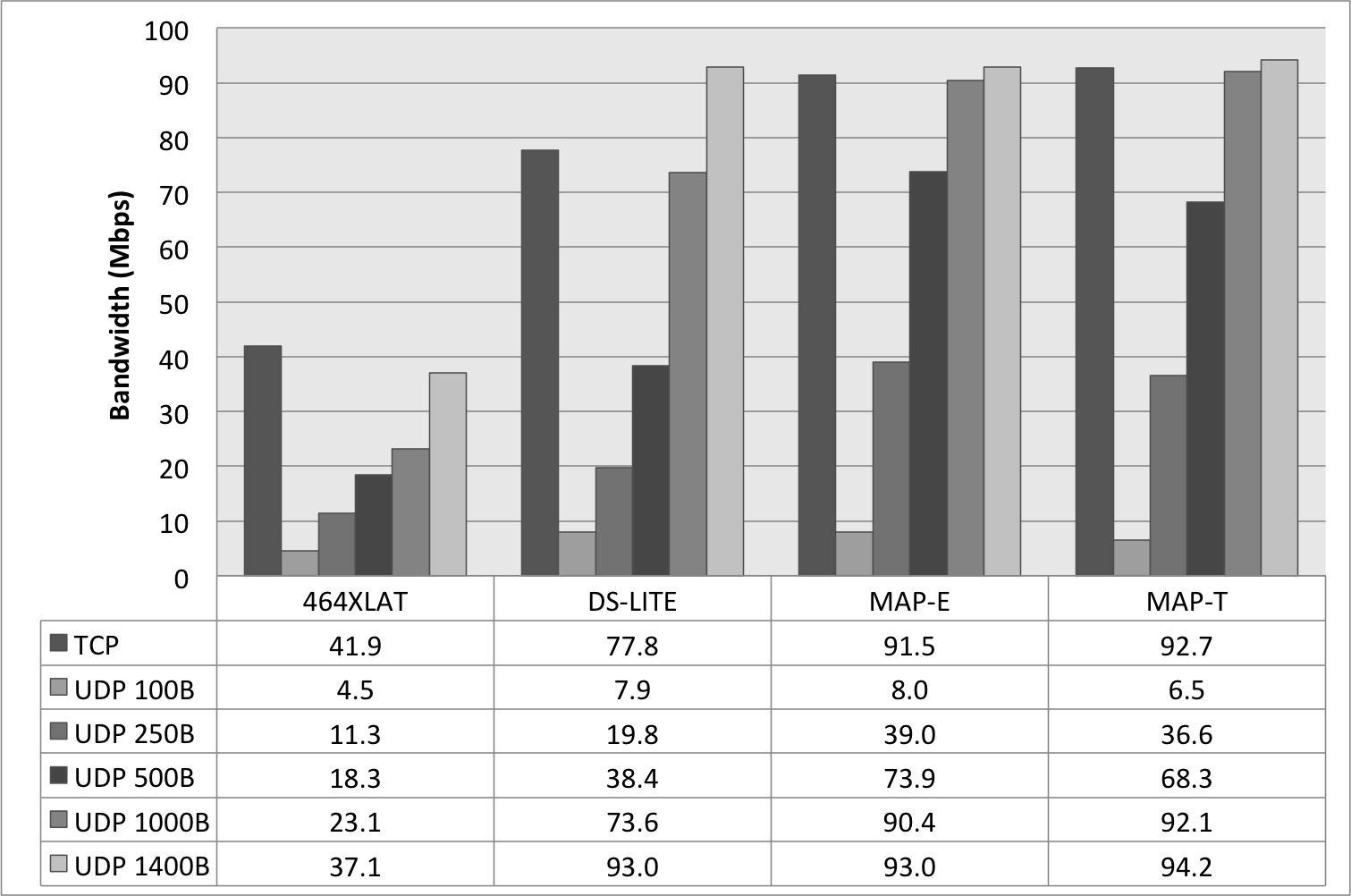}
  \caption[Throughput x Packet Size x Technique] {Throughput x Packet Size x Technique\label{bXtpXt}}
\end{figure}

Because of the packet loss the real UDP throughput was always smaller than the 100 Mbps generated, as showed on figure \ref{bXtpXt}. For UDP packets of 1000 or 1400 bytes MAP-T and MAP-E results were similar to the TCP performance, but as the packet loss was very big for smaller packets the throughput was also smaller. DS-Lite had an increasing throughput results as the packet sizes were getting bigger and had similar performance to MAP-E and MAP-T for the bigger UDP packets, despite being slower on the TCP test. 464XLAT had the worse performance in all tests with a maximum throughput of 40 Mbps on the TCP test.

The encapsulation techniques, DS-Lite and MAP-E, added a 40 bytes overhead for each packet. This overhead was the IPv6 header added to the IPv4 packet to allow the packet to go through the IPv6 only network. It means a 40\% overhead for the smaller UDP packet, but only a 2.8\% overhead for the bigger packet. For the double translation techniques, 464XLAT and MAP-T, the added size to each packet was 20 bytes, half of the value of the encapsulation. It happens because the original 20 bytes IPv4 header is not transmitted on the IPv6 only network, but it is replaced by the 20 bytes bigger IPv6 header. Considering those results the double translation techniques would be expected to have better performance, but as the IPv6 only network was a Gigabit Ethernet, ten times bigger than the 100 Mbps used on the UDP tests, the lower overhead wasn't a advantage for the double translation techniques. 

\section{Conclusion}
\label{conclusion}

As mentioned on the Network Topology section, the 464XLAT doesn't have a CPE dedicated open source implementation, as DS-Lite, MAP-E and MAP-T and the software used to implement it was last updated on June 2011. Because of that the translation used almost all CPU for every packet size and still were unable to achieve 40 Mbps with the biggest UDP packets. As the 464XLAT is currently being used at T-Mobile's production network the bad result obtained is probably related to the software implementation not to the technique. 

On the DS-Lite, MAP-E and MAP-T networks the low speed obtained for the smaller packets seems directly related to a packets per second limitation on the CPE, as the packet loss for the smaller packets was bigger than 80\%, while the packet loss was smaller than 8\% for the 1400 bytes packets.

The DS-Lite CPU usage at the CPE is almost constant for all the packet sizes and this high CPU usage, as it also happened in the 464XLAT, is the reason for the two times bigger latency than on MAP-E and MAP-T networks.

By comparing encapsulation versus double translation of MAP-E against MAP-T, as both were implemented for the same group and share most of the source code, it seems that encapsulation is less processor intensive and this trend clearly visible when using bigger packets. The smaller processor usage also reflect on MAP-E having a higher throughput on the small packets.

All techniques proved to be a reasonable alternative for the IPv4 to IPv6 transition on an IPv6 only network, but the available open source must be optimized or a CPE with higher computational power must be used to enable large scale deployments with any of the techniques analyzed in this comparison.


\begin{backmatter}

\section*{Competing interests}
The authors declare that they have no competing interests.

\section*{Author's contributions}
EC designed and built the test networks and defined the tests, made some data acquisition, analyzed and interpreted the data and drafted the manuscript. RC made some data acquisition and preliminary data analysis and revision of the manuscript. WZ contributed to the conception of the test networks, definition of the tests, reviewed and have given final approval to publish the manuscript.

\section*{Acknowledgements}
We would like to acknowledge the support from NIC.br (Brazilian Network Information Center) that borrowed the necessary equipment and provided the network access to the development of the tests. 


\bibliographystyle{bmc-mathphys} 
\bibliography{ComparisonBetweenIPv4toIPv6TransitionTechniques}      

\end{backmatter}
\begin{figure}[b]
\centering
\includegraphics[keepaspectratio=true]{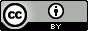}

Comparison Between IPv4 to IPv6 Transition Techniques by Edwin Cordeiro is licensed under a Creative Commons Attribution 4.0 International License. 
\end{figure}
\end{document}